\documentclass[12pt]{article}
\usepackage[mathscr]{eucal}
\usepackage{amssymb,latexsym}
\usepackage{verbatim}
\usepackage{amsmath}
\usepackage{amsthm}
\usepackage{enumerate}
\usepackage{authblk}
\usepackage{color}
\usepackage{url}

\usepackage{tikz}
\usetikzlibrary{matrix}
\usetikzlibrary{snakes}
\usetikzlibrary{arrows,calc,shapes,decorations.pathreplacing}
\usepackage{tikz-cd}

\usepackage[normalem]{ulem}

\usepackage{amscd,amssymb,amsthm,verbatim}
\usepackage{enumerate}
\usepackage{bm}
\usepackage{mathrsfs, graphicx} 
\usepackage{stmaryrd}

\usepackage{setspace}

\usepackage[%
bookmarks=true,
colorlinks,
linkcolor=blue,
urlcolor=blue,
citecolor=blue,
plainpages=false,
pdfpagelabels,
final,
breaklinks=true\between
]{hyperref}
\usepackage{hyperref}

\newtheorem{thm}{Theorem}
\newtheorem{lem}[thm]{Lemma}
\newtheorem{prop}[thm]{Proposition}
\newtheorem{cor}[thm]{Corollary}

\renewcommand\l{\lambda}

\newcommand\wt{\widetilde}

\renewcommand\S{\Sigma}

\newcommand\e{\varepsilon}

\renewcommand\div{{\rm div}}

\newcommand\<{\langle}
\renewcommand\>{\rangle}

\renewcommand\l{\lambda}
\newcommand\g{\gamma}

\newcommand\beq{\begin{equation}}
\newcommand\eeq{\end{equation}}
\newcommand\ben{\begin{enumerate}}
\newcommand\een{\end{enumerate}}
\newcommand\bit{\begin{itemize}}
\newcommand\eit{\end{itemize}}

\DeclareMathOperator{\diver}{div}

\renewcommand{\div}{\diver}

\newcommand{\tr}{\mathrm{tr}\,}

\newcommand{\R}{\mathbb R}

\newcommand{\ov}{\overline}

\newcommand{\pd}{\partial}

\newcommand{\Z}{\mathbb{Z}}

\newcounter{mnotecount}

\title{Rigidity aspects of Penrose's singularity theorem}

\author[1]{Gregory Galloway\footnote{galloway@math.miami.edu}}
\author[2]{Eric Ling\footnote{el@math.ku.dk}}

\affil[1]{University of Miami, Coral Gables, FL, USA}

\affil[2]{Copenhagen Centre for Geometry and Topology (GeoTop),
\linebreak
Department of Mathematical Sciences, \linebreak University of Copenhagen, Denmark}

\begin{document}
\date{}
\maketitle

\vspace{.15in}

\begin{abstract} 
 In this paper,  we study rigidity aspects of Penrose's singularity theorem. Specifically, we aim to answer the following question: if a spacetime satisfies the hypotheses of Penrose's singularity theorem except with weakly trapped surfaces instead of trapped surfaces, then what can be said about the global spacetime structure if the spacetime is null geodesically complete? In this setting, we show that we obtain a foliation of MOTS which generate totally geodesic null hypersurfaces. Depending on our starting assumptions, we obtain either local or global rigidity results. We apply our arguments to cosmological spacetimes (i.e., spacetimes with compact Cauchy surfaces) and scenarios involving topological censorship.
\end{abstract}

\begin{spacing}{.25}
\tableofcontents 
\end{spacing}

\section{Introduction}
Penrose's original 1965 singularity theorem \cite{Penrose1965}, which earned him half of the 2020 Nobel Prize, predicts an incomplete null geodesic (i.e., the singularity) emanating from a closed trapped surface $\S$, provided the spacetime admits a noncompact Cauchy surface and satisfies the null energy condition. Closed trapped surfaces are believed to form during immense gravitational collapse where the spacetime curvature is so strong that, in a vague sense, light cannot escape the gravitational field. Mathematically, they occur when the inner null expansion $\theta^-$ and the outer null expansion $\theta^+
$ are both negative. In a suitably defined black hole spacetime, closed trapped surfaces must lie within the event horizon.

Focusing solely on the outer direction, a closed surface $\S$ is outer trapped if $\theta^+ < 0$, weakly outer trapped if $\theta^+\leq 0$, and marginally outer trapped if $\theta^+ = 0$; in the last case, $\S$ is referred to as a marginally outer trapped surface (MOTS). For stationary black hole spacetimes, MOTS occur as spatial cross sections of the event horizon. In general dynamical spacetimes, MOTS typically lie within the event horizon, where they can form the boundary of the trapped region.

In this paper, we investigate the rigidity aspects of Penrose's singularity theorem. Ignoring some technicalities, the question we aim to answer is the following:

\medskip

 \noindent \emph{If a spacetime satisfies the hypotheses of Penrose's singularity theorem except that $\S$ is only weakly outer trapped, then what can be said about the global structure of the spacetime if we assume that it is null geodesically complete? }

\medskip

 In section \ref{sec: local rig}, we establish a local rigidity result: If the spacetime is null geodesically complete, then $\S$ is a MOTS and there is a neighborhood of MOTS around $\S$ in the Cauchy surface whose outward null normal geodesics generate totally geodesic null hypersurfaces; see Theorem \ref{thm: local rigidity 1}. If $\S$ is also weakly inner trapped, then a spacetime neighborhood of $\S$ is foliated by totally geodesic codimension 2 spacelike submanifolds, each diffeomorphic to $\S$; see Theorem \ref{thm: local rigidity 2}. The local foliation by MOTS is obtained via an inverse function theorem argument as in, e.g., \cite{GalMots1, GalMots2, EGM, GalMendes, Mendes}. The existence of the totally geodesic null hypersurface is proved in \cite{EGP} and follows from the null maximum principle \cite{Gnullmp}.

 Moving to more global considerations, in section \ref{sec: cpt_with_bdy}, we establish rigidity results in the case where the Cauchy surface contains a compact, connected top-dimensional submanifold with two boundary components, $\S$ and $S$, which are weakly outer trapped and weakly outer untrapped, respectively. In this case, null geodesic completeness implies that the region between $\S$ and $S$ is foliated by MOTS and each generates a totally geodesic null hypersurface; see Theorem \ref{thm: glob rigidity 1}. Our proof relies on a continuity argument used in \cite{EGM}.
 
In section \ref{sec: cosmo}, we apply these results to cosmological spacetimes, i.e., spacetimes with compact Cauchy surfaces, when $\S$ is a nonseparating surface. In this case, we obtain rigidity information on the whole spacetime.
 In section~\ref{sec: top cen}, we obtain rigidity results within the context of topological censorship.  When $\text{dim}(M) = 4$ and $\S$ is topologically a two-sphere, our results are decisive, see Corollary \ref{cor: top cen}.

Our results can be paraphrased in the following way ``in the presence of a weakly trapped surface, either the spacetime is null geodesically complete (future and past) or the spacetime is very special in a suitably rigid sense." Penrose's original singularity theorem concluded only future null incompleteness, and our results can be interpreted in this spirit; see specifically Proposition \ref{prop: totally geodesic} in section \ref{sec: prelim}. Moreover, sections \ref{sec: cpt_with_bdy} and \ref{sec: top cen} are consistent with the setting of gravitational collapse developing from a nonsingular past. In this setting, our results imply that either the spacetime is future null geodesically incomplete or ``the spacetime is very special in a suitably rigid sense." A similar type of interpretation holds in the cosmological setting (section \ref{sec: cosmo}) where one naturally assumes future null completeness. In this setting, our results imply that either the spacetime is past null geodesically incomplete or ``the spacetime is very special in a suitably rigid sense."

\section{Preliminaries}\label{sec: prelim}

Let $(M^{n+1},g)$ be a globally hyperbolic spacetime, $n \geq 2$. An \emph{initial data set} $(V^n,h,K)$ for $(M,g)$ is a spacelike Cauchy hypersurface $V \subset M$ with induced metric $h$ and second fundamental form $K$. Let $\Sigma^{n-1}$ be a connected and compact two-sided hypersurface in $V$. Let $\nu$ be one of the two unit normal vector fields on $\Sigma$ pointing into $V$. We call $\nu$ the `outward' direction and $-\nu$ the `inward' direction. Let $u$ be the future directed unit normal on $V$. $\S$ admits two future directed null normal vector fields, $\ell^{\pm} = u \pm \nu$. The \emph{null second fundamental forms} $\chi^{\pm}$ and \emph{null expansion scalars} $\theta^{\pm}$ are given by 
\[
\chi^{\pm}(X,Y) \,=\, \<\nabla_X \ell^{\pm}, Y\> 
\]
for all $X,Y \in T_p\S$ and $p \in \S$, and
\[
\theta^{\pm} \,=\, \tr_\S \chi^{\pm} \,=\, \div_\S \ell^{\pm} \,=\, \tr_\S K \pm H,
\]
where $H = \div_\S \nu$ is the mean curvature of $\Sigma$ within $V$.

$\S$ is \emph{outer trapped} if $\theta^+ < 0$, \emph{weakly outer trapped} if $\theta^+ \leq 0$, and \emph{marginally outer trapped} if $\theta^+ = 0$. In the latter case, $\S$ is referred to as a \emph{marginally outer trapped surface} (MOTS). Evidently, in the time-symmetric setting ($K = 0$), a MOTS is a minimal surface in $V$. Analogous terminology holds for \emph{MITS},  $\theta^- = 0$.

Let $\S \subset V$ be a MOTS, and let $\{\S_t\}_{|t| < \e}$ be a variation of $\S$ determined by $\S_t \colon x \mapsto \exp_x\big(t \phi(x)\nu(x)\big)$ for some $\phi \in C^\infty(\Sigma)$. Let $\theta^+(t)$ denote the outward null expansion of $\S_t$. From e.g. \cite[Eq.\ 1]{AMSPRL} (see also \cite[Prop.\ 7.32]{DLee}), we have that
\[
\frac{\pd \theta^+}{\pd t}\bigg|_{t=0} \,=\, L(\phi), 
\]
where
\[
L(\phi)\,=\,-\Delta \phi + 2 \<X, \nabla \phi\> + (Q + \div X - |X|^2)\phi.
\]
Here $X$ is the tangential part of $\nabla_\nu u$ on $\S$  and
\[
Q \,=\, \frac{1}{2}S - G(u, \ell^+) - \frac{1}{2}|\chi^+|^2,
\]
where $S$ is the scalar curvature on $\S$ and $G$ is the Einstein tensor for $M$. We remark that, in terms of initial data $(V,h,K)$, $X$ is metrically dual to the one-form $K(\nu, \cdot)|_\Sigma$.

In the time-symmetric case ($K=0$), $\theta^+ = H$, $X = 0$, and $L$ reduces to the classical stability operator of minimal surface theory. In analogy with the minimal surface case, we refer to $L$ as the \emph{MOTS stability operator}. Note, however, that $L$ is not in general self-adjoint. 

Nevertheless, it is shown in \cite{AMSPRL,AMS08} (see also 
\cite[Thm. A.10]{DLee}) that, as a consequence of the Krein-Rutman theorem, $L$ admits a {\it principal eigenvalue} $\l_1$, which is real and simple. Furthermore there exists an associated principal eigenfunction $\phi$ which can be chosen to be strictly positive.  
Moreover, $\l_1 \geq 0$ (respectively, $\l_1 > 0$) if and only if there exists a  positive $\psi \in C^\infty(\Sigma)$  such that $L(\psi) \geq 0$ (respectively, $L(\psi) > 0$). 
In analogy with the minimal surface case, we say a MOTS $\S$ is \emph{stable} if $\l_1 \geq 0$.

The following lemma has been used in various rigidity results for MOTS; see \cite{GalMots1, GalMots2, EGM, GalMendes, Mendes}.
As it does not appear as a stand-alone result in the literature, we provide a proof in the appendix for the convenience of the reader.

\medskip

\begin{lem}\label{lem: IVT argument}
Let $\S$ be a stable MOTS in an initial data set $(V,h,K)$.
If $\l_1 = 0$, then there is a neighborhood $U$ of $\S$ in $V$ such that, up to isometry,
\[
U \,=\, (-\e, \e) \times \S \quad \text{ and } \quad h|_U \,=\, \phi^2 dt^2 + \g_t,
\]
where $\phi = \phi(t,x)$ is a smooth function on $U$ and $\g_t$ is the induced metric on $\S_t = \{t\} \times \S$. Moreover, the outward null expansion of each $\S_t$ is constant, i.e., $\theta^+(t)$ is  constant on $\S_t$ where $\theta^+(t)$ is computed with respect to $\ell^+(t) = u + \nu(t)$, where $\nu(t) = \frac{1}{\phi(t, \cdot)}\pd_t$ is the outward unit normal to $\S_t$.
\end{lem}

The next theorem can be thought of as a `one-sided version' of Penrose's singularity theorem, and it's proved in a similar way, see e.g.,  \cite[Thm. 7.1]{AMMS}.

\begin{thm}\label{thm: one-sided Penrose}
Let $(M,g)$ be a globally hyperbolic spacetime satisfying the null energy condition with a spacelike Cauchy hypersurface $V$. Let $\Sigma$ be a connected, closed hypersurface in $V$ which separates $V$. If $\S$ is outer trapped (i.e., $\theta^+ < 0$) with respect to a component of $V \setminus \S$ whose closure in $V$ is noncompact, then $M$ is future null geodesically incomplete.
\end{thm}

The following proposition is a rigid version of the one-sided Penrose singularity theorem.

\begin{prop}\label{prop: totally geodesic}
Let $(M,g)$ be a globally hyperbolic spacetime satisfying the null energy condition with a spacelike Cauchy hypersurface $V$. Let $\S$ be a connected, closed hypersurface in $V$ which separates $V$. Suppose $\S$ is a weakly trapped surface (i.e., $\theta^+ \leq 0$) with respect to the future directed null normal $\ell^+$ which points toward a component of $V \setminus \S$ whose closure is noncompact. 
Then either $M$ is future null geodesically incomplete or else $\Sigma$ is a MOTS (in fact $\chi^+ =0$) and the future inextendible null normal geodesics in the direction of $\ell^+$ form an achronal totally geodesic\footnote{By a totally geodesic null hypersurface, we mean that it's null second fundamental form vanishes everywhere and has the usual interpretation that geodesics tangent to it remain within it.} null hypersurface.
\end{prop}

Proposition \ref{prop: totally geodesic} is proved in the appendix of \cite{EGP} in the case that $\Sigma$ is a MOTS ($\theta^+ = 0)$, but the proof trivially carries over to the case when $\Sigma$ is weakly trapped.

\section{Local rigidity}\label{sec: local rig}

In this section, we present two Penrose-type  local rigidity results.

\begin{thm}\label{thm: local rigidity 1}
Let $(M,g)$ be a globally hyperbolic spacetime satisfying the null energy condition with a spacelike Cauchy hypersurface $V$. Let $\S$ be a connected, closed hypersurface in $V$ that separates $V$ into two components, both having noncompact closures in $V$. Suppose $\S$ is weakly outer trapped, $\theta^+ \leq 0$, with respect to one of the components of $V \setminus \S$. Then either
\begin{itemize}
\item[\emph{(a)}] $M$ is null geodesically incomplete, or else,
\item[\emph{(b)}] there is a neighborhood $U$ of $\S$ in $V$ diffeomorphic to $(-\e, \e) \times \S$ such that for each $t \in (-\e, \e)$,  $\S_t = \{t\} \times \S$ is a MOTS (in fact $\chi^+_t = 0)$ and the  outward null normal geodesics to $\S_t$ 
(past and future complete)
 form a closed achronal totally geodesic null hypersurface $N_t$.
\end{itemize} 
\end{thm}

\proof
Suppose $M$ is null geodesically complete. If there is a point on $\S$ where $\theta^+ < 0$, then, by a result of Andersson and Metzger \cite[Lem. 5.2]{AndMetz}, $\S$ can be deformed via null mean curvature flow within $V$ to an outer trapped surface $\S'$ which contradicts null geodesic completeness by Theorem \ref{thm: one-sided Penrose}. Thus $\S$ is a MOTS.

Let $\l_1$ be the principal eigenvalue of the MOTS stability operator $L$ with respect to $\S$. We show $\l_1 = 0$. Let $\phi$ be the associated positive eigenfunction and consider the variation $\S_s \colon x \mapsto \exp_x \big(s\phi(x)\nu(x)\big)$. Let $\theta^+(s)$ denote the corresponding outward null expansions. Then
\[
\frac{\pd \theta^+}{\pd s}\bigg|_{s = 0} \,=\, L(\phi) \,=\, \l_1\phi.
\]
If $\l_1 < 0$, then $\theta^+(s) < 0$ for small $s > 0$, and hence there are outer trapped surfaces in $V$ to the outside of $\S$, contradicting future null completeness of $M$. Similarly, if $\l_1 > 0$, then $\theta^+(s) < 0$ for small $s < 0$, again contradicting future null completeness. Thus $\l_1 = 0$.

By Lemma \ref{lem: IVT argument}, there exists a neighborhood $U$ of $\S$ in $V$, with $U \approx (-\e, \e) \times \S$, such that $\S_t = \{t\} \times \S$ has constant outward null expansion $\theta^+(t)$. We must have $\theta^+(t) \geq 0$, otherwise we contradict future completeness. Suppose $\theta^+(t) > 0$. Note that
\[
\theta^+(t) \,=\, \div_{\S_t}\big(u + \nu(t)\big) \,=\, -\div_{\S_t}\big(- u - \nu(t)\big).
\]
Therefore $\S_t$ is inner trapped if we reverse the time orientation. Since the component in the inward direction also has noncompact closure, this contradicts past null completeness. Thus $\theta^+(t) = 0$ for all $t \in (-\e,\e)$, i.e., $\S_t$ a MOTS for all $t$. Proposition~\ref{prop: totally geodesic} along with its time-dual produce two totally geodesic achronal null hypersurfaces with boundary given by $\S_t$; one is formed to the future, call it $F_t$, and the other formed to the past, call it $P_t$. We glue $F_t$ and $P_t$ together along $\S_t$ forming a closed, totally geodesic null hypersurface $N_t$. Achronality of $N_t$ follows since $\S_t$ separates $V$. Indeed, if $\g$ is a timelike curve starting on $P_t$ and ending on $F_t$, then the integral curves of a timelike vector field project $\g$ down to a curve in $V$ which must intersect $\S_t$ at some point $x$. By concatenating this integral curve from $x$ with a portion of $\g$, we obtain a contradiction either with the achronality of $F_t$ or the achronality of $P_t$. 
\qed

\medskip
\medskip

\begin{thm}\label{thm: local rigidity 2}
Let $(M,g)$ be a globally hyperbolic spacetime satisfying the null energy condition with a spacelike Cauchy hypersurface $V$. Let $\S$ be a connected, closed hypersurface in $V$ that separates $V$ into two components, both having noncompact closures in $V$. Suppose $\S$ is weakly trapped, i.e., $\theta^+ \leq 0$ and $\theta^- \leq 0$  with respect to one of the unit normals. Then either
\begin{itemize}
\item[\emph{(a)}] $M$ is null geodesically incomplete, or else,

\item[\emph{(b)}] $\theta^\pm = 0$ and there are neighborhoods $U^\pm  \approx (-\e^\pm, \e^\pm) \times \S$ of $\S$ in $V$ such that $U^+$ and $U^-$ are foliated by MOTS (in fact $\chi^+ = 0$) and MITS (in fact $\chi^- = 0$), respectively. Each MOTS, $\S^+_t$, generates a closed achronal totally geodesic null hypersurface, $N^+_t$, via the generators of the outward null normal $\ell^+_t$. Likewise, each MITS, $\S^-_s$, generates an analogous $N^-_s$. Moreover, the intersection of the two families, $N^+_t$ and $N^-_s$, foliate a spacetime neighborhood of $\S$ by totally geodesic codimension \emph{2} spacelike submanifolds, each diffeomorphic to $\S$.

\end{itemize}
\end{thm}

\proof
Suppose $M$ is null geodesically complete. We are assuming $\S$ is 
weakly outer and inner trapped with respect to one of the components of $V \setminus \S$. As in the proof of Theorem \ref{thm: local rigidity 1}, $\S$ is both a MOTS and a MITS. There are neighborhoods $U^+$ and $U^-$ of $\S$ in $V$ foliated by MOTS and MITS, respectively. Call these foliations $U^+ = \{\S_t^+\}$ and $U^- = \{\S_s^-\}$ for $t \in (-\e^+, \e^+)$ and $s \in (-\e^-, \e^-)$, and note that they coincide at $\S = \S_0^+ = \S_0^-$.  Each $\S_t^+$ generates a totally geodesic outward null hypersurface $N^+_t$, i.e., it has vanishing null second fundamental. Likewise with the inner direction on $N_s^-$.

Let $N$ be the union of the intersections  $\{N^+_t \cap N^-_s \mid t \in (-\e^+, \e^+),\: s \in (-\e^-, \e^-)\}$.  Then $N$ is nonempty since $\S \subset N$. Since the null generators of these null hypersurfaces are distinct, it follows that $N^+_t$ and $N^-_s$ must intersect transversely (if they do at all). It follows that $N$ is an open set in $M$ containing $\S$. Clearly, since $N^+_0 \cap N^-_0 = \S$, $N^+_t \cap N^-_s$ will intersect in a spacetime neighborhood of $\S$ for all $t$ and $s$ sufficiently small in magnitude.
 The intersections within this  neighborhood are codimension 2 compact submanifolds which are totally geodesic since it's the intersection of two transverse totally geodesic null hypersurfaces. The null geodesic generators of $N^+_t$ establish a diffeomorphism from each intersection $N^+_t \cap N^-_s$ to $\S^+_t$ and hence they are diffeomorphic to $\S$.  Indeed, by an achronality argument, the null geodesic generators of $N^+_t$ intersect $N^-_s$ only once. Therefore the exponential map sets up a well-defined embedding of $\Sigma^+_t$ into $N^+_t \cap N^-_s$ producing the diffeomorphism.
\qed

\medskip
\medskip

\noindent\emph{Remark.} For $t < 0$ and $s > 0$,  the intersection $N^+_t \cap N^-_s$ lies in $J^+(\Sigma)$. To see this, let $V^\pm$ denote the connected components of $V \setminus \S$ with $\nu|_\Sigma$ pointing into $V^+$. Since $\S$ separates, causal-theoretic arguments as in the proof of Theorem \ref{thm: local rigidity 1} show that $J^+(\ov{V^-}) = J^+(\Sigma) \sqcup D^+(V^-)$, where $D^+$ is the future domain of dependence. The corresponding property holds for $V^+$ as well. Therefore, $J^+(\ov{V^-}) \cap J^+(\ov{V^+}) = J^+(\Sigma)$. 

\section{Compact-with-boundary rigidity}\label{sec: cpt_with_bdy}

It is often assumed that  the process of gravitational collapse and formation of a black hole develops from a nonsingular past. In this section, we will take this point of view. 

\medskip
\medskip

\begin{thm}\label{thm: glob rigidity 1}
Let $(M,g)$ be a past null geodesically complete globally hyperbolic spacetime satisfying the null energy condition with a spacelike Cauchy hypersurface $V$. Let $\Sigma$ be a connected, closed hypersurface in $V$ that separates $V$ into two components, both having noncompact closures in $V$. Let $W$ be a compact, connected top-dimensional submanifold in $V$ with boundary $\pd W = \S \sqcup S$. ($S$ may have multiple components.) Assume that $\theta^+_S \geq 0$ with respect to the normal pointing out of $W$.

If $\S$ is weakly outer trapped,  $\theta^+_\Sigma \leq 0$, with respect to the normal pointing into $W$, then either
\begin{itemize}
\item[\emph{(a)}] $M$ is future null geodesically incomplete, or else,
\item[\emph{(b)}] $S$ is diffeomorphic to $\S$ and $W$ is diffeomorphic to $[0,\delta] \times \S$. Moreover, for each $t \in [0,\delta]$, $\S_t = \{t\} \times \S$ is a MOTS (in fact $\chi^+_t = 0$) and the outward null normal geodesics to $\S_t$ form a closed achronal totally geodesic  null hypersurface $N_t$.  
\end{itemize}
\end{thm}

\proof
This follows from a continuity argument as in the proof of Theorem 1.2 in \cite{EGM}. Assume $M$ is future null geodesically complete. Briefly, one shows that the leaves of the local foliation $\{\S_t\}_{t \in [0,\e)}$ from Theorem \ref{thm: local rigidity 1} have a smooth embedded limit $\S_\e$. (This uses the fact that, since $\chi_t^+= 0$, the second fundamental forms of the $\S_t$'s within $W$ are uniformly bounded in terms of the second fundamental form of $W$ in $M$.) Then the strong maximum principle (\cite[Prop. 3.1]{AshtekarGalloway}, \cite[Prop. 2.4]{AndMetz}) shows that $\S_\e = S$ if $\S_\e \cap S \neq \emptyset$. If $\S_\e \cap S = \emptyset$, then we can continue the foliation. Connectedness of $W$ implies $W 
\approx [0, \delta] \times \S$ with $S \approx \Sigma_\delta$.
\qed

\medskip
\medskip

\noindent\emph{Remark.}   Suppose in Theorem \ref{thm: glob rigidity 1}, one assumes that $V$ is geodesically complete, with bounded geometry, in the sense described in \cite[Sec.\ 1]{MSY} (called `homogeneous regularity' there). Suppose further that $K$ is uniformly bounded on all of $V$.   Now let $W$ be the closure of one of the two components separating $V$. (Note that $W$ in this case is noncompact.)  One can again argue that the leaves of the local foliation 
$\{\S_t\}_{t \in [0,\e)}$ from Theorem \ref{thm: local rigidity 1} have a smooth embedded compact limit 
$\S_\e$ provided the union of the leaves has a limit point in $W \setminus \S$. Then, again by a continuation argument, if $M$ is future null geodesically complete, the conclusions~(b) of Theorem \ref{thm: glob rigidity 1} hold, with $[0, \delta]$ replaced by $[0, \delta)$,  $\delta \in (0, \infty]$.  We thank Michael Eichmair for clarifications regarding this.

\medskip
\medskip

In the following, we apply the previous theorem to a situation where $\Sigma$ is both weakly outer and inner trapped. A priori, the null expansion assumptions on $S$ are compatible with an asymptotically flat end.

\medskip

\begin{thm}\label{thm: glob rigidity 2}
Let $(M,g)$ be a past null geodesically complete globally hyperbolic spacetime satisfying the null energy condition with a spacelike Cauchy hypersurface $V$. Let $\Sigma$ be a connected and closed hypersurface in $V$ that separates $V$ into two components, both having noncompact closures in $V$. Let $W$ be a compact, connected top-dimensional submanifold in $V$ with boundary $\pd W = \S \sqcup S$. Assume $S$ is weakly outer untrapped and weakly inner trapped, i.e., $\theta^+_S \geq 0$ and $\theta^-_S \leq 0$, with respect to the normal pointing out of $W$.

If $\S$ is weakly trapped,  $\theta^\pm_\Sigma \leq 0$, with respect to the normal pointing into $W$, then either
\begin{itemize}
\item[\emph{(a)}] $M$ is future null geodesically incomplete, or else,
\item[\emph{(b)}] $W$ is foliated by a family of MOTS, $\Sigma^+_t$, and a family of MITS, $\Sigma^-_s$, each generating totally geodesic null hypersurfaces $N^+_t$ and $N^-_s$ as in Theorem \ref{thm: local rigidity 2}. The two foliations of null hypersurfaces together cover $J^+(W) \cup J^-(W)$. Moreover, the intersection of the two families of null hypersurfaces cover the domain of dependence $D(W) = D^+(W) \cup D^-(W)$ by totally geodesic codimension \emph{2} spacelike submanifolds.
\end{itemize}
\end{thm}

\proof
Assume $M$ is future null geodesically complete. Theorem \ref{thm: local rigidity 2} implies that $\theta^\pm_\S = 0$. Theorem \ref{thm: glob rigidity 1} gives the existence of the foliating family of MOTS, $\{\S^+_t\}_{t \in [0, \delta^+]}$, and MITS, $\{\S^-_s\}_{s \in [0, \delta^-]}$ with $\S^+_0 = \S^-_0 =\S$.  Let $N^+_t$ and $N^-_s$ denote the corresponding closed achronal totally geodesic null hypersurfaces.

Consider the collection of unions  $\{N^+_t \cup N^-_s \mid t \in [0, \delta^+],\: s \in [0, \delta^-]\}$. Let $A$ denote their union. We want to show $A = J^+(W) \cup J^-(W)$. Clearly $A$ is a subset. It suffices to show $A$ is both open and closed in the subspace topology. Fix $p \in A$ and suppose without loss of generality that $p \in N^+_t$ for some $t \in [0, \delta^+]$. If $t \in (0,\delta^+)$, then the exponential map produces a neighborhood of points around $p$ all coming from $N^+_{t'}$ for $t' \in (t - \e, t +\e)$. In the case that $t = 0$ or $t = \delta^+$, we can extend the foliation a little further to generate a neighborhood of points arising from the extended foliation and intersect this neighborhood with $J^+(W) \cup J^-(W)$ producing a relative neighborhood in the subspace topology. Thus $A$ is open in $J^+(W) \cup J^-(W)$. To see that $A$ is closed, let $q_n \in A$ be a sequence limiting to $q$. Without loss of generality, we can assume $q_n \in N^+_{t_n}$ for $t_n \in [0, \delta^+]$. Each $q_n$ arises from some $p_n \in \Sigma_{t_n}$ via $q_n = \exp_{p_n}(k_n)$, where $k_n$ is an outward normal null vector. A subsequence of $p_n$ converges to $p \in \S_t$ for some $t$. Let $q' \in I^+(q)$, then $J^-(q') \cap I^+(V)$ is a compact set which contains all but finitely many $q_n$. Strong causality implies that a subsequence of $k_n$ converges to some null vector $k \in T_pM$ or converges to $k = 0$. In the latter case, we're done. In the former case, continuity shows that $k$ is an outward normal null vector and $\exp_p(k) = q$. 

Let $N$ denote the union of the intersections $\{N^+_t \cap N^-_s \mid t \in [0, \delta^+],\: s \in [0, \delta^-]\}$. It suffices to show $N$ is a subset of $D(W)$, for then a similar open and closed argument as above establishes 
$N = D(W)$.
Fix $q \in N^+_t \cap N^-_s$ and without loss of generality, assume $q \in I^+(W)$. We want to show $q \in D^+(W)$. There is a $p_+ \in \S^+_t$ and $p_- \in \S^-_s$ such that there are outward/inward null normal geodesics from $p_+$/$p_-$ to $q$. The separating property of the $\S$'s implies that $p_+$ is on the inside of $\S^-_s$ and $p_-$ is on the outside of $\S^+_t$. It follows that the closed region \emph{between} $\S^+_t$ and $\S^-_s$ within $V$ is nonempty. Call this region $R$. Let $\g$ be a past-directed and past-inextendible causal curve starting from $q$. It suffices to show that $\g$ intersects $R$. Clearly there is a timelike curve $\l$ from $R$ to $q$. Let $X$ be a global timelike vector field such that $\l$ is an integral curve of $X$. $\g$ meets $V$ at some point $p$. If $\g$ did not intersect $R$, then, $p$ lies to the inside of $\S^+_t$ or to the outside of $\S^-_s$. Assume the former. 
The timelike integral curves of $X$ project $\g$ down to $V$ and generate a curve from $p$ to the starting point of $\l$; this curve must intersect $\S^+_t$. Hence there is a timelike curve from 
$\S^+_t$ to $q$, which contradicts the achronality of $N^+_t$.\qed

\medskip
\medskip

\noindent\emph{Remark.} 
The second fundamental form $K$ on $V$ is said to be \emph{$(n-1)$-convex},  if, at every point, the sum of the smallest $n-1$ eigenvalues of $K$ with respect to the induced metric $g|_V$ is nonnegative.  It can be seen that this implies $\tr_\Sigma K\ge0$ for every hypersurface $\Sigma\subset V$.  If we assume in Theorem \ref{thm: glob rigidity 2} that $K$ is $(n-1)$-convex, then the two foliations of MOTS $\S^+_t$ and MITS $\S^-_s$ actually agree with each other. To see this, fix $\Sigma^-_s$. We have $\theta^-_s = 0$; hence $H = \text{tr} K \geq 0$
 on $\Sigma^-_s$ by $(n-1)$-convexity. Therefore $\theta^+_s \geq 0$; hence $\theta^+_s = 0$. By compactness, there is a $t$ such that $\S^+_t$ and $\S^-_s$ are tangent and $\S^-_s$ lies to the inside of $\S^+_t$. Since $\theta^+_t = \theta^-_s = 0$, it follows that $\S^-_s = \S^+_t$ by the maximum principle for $\theta^+$ \cite[Prop. 3.1]{AshtekarGalloway}.

\section{Cosmological rigidity}\label{sec: cosmo}

In the cosmological context (i.e., spacetimes with compact Cauchy surfaces), we have the following two theorems.

\medskip

\begin{thm}\label{thm: cpt with boundary 1}
Let $(M,g)$ be a future null geodesically complete globally hyperbolic spacetime satisfying the null energy condition. Let $V$ be a spacelike compact Cauchy hypersurface, and let $\S$ be a closed nonseparating two-sided hypersurface in $V$, which is weakly outer untrapped, i.e., $\theta^+ \geq 0$. Then either
\begin{itemize}
\item[\emph{(a)}] $M$ is past null geodesically incomplete, or else,
\item[\emph{(b)}] $M$ is foliated by closed totally geodesic null hypersurfaces.
\end{itemize}
\end{thm}

\proof
Assume $M$ is past null geodesically complete. Since $\S$ is nonseparating, $M$ admits a covering spacetime $\wt{M}$  with a Cauchy surface $\wt{V}$ which has $\Z$ isometric copies of $\S$, call them $\wt{\S}_n$, such that each $\wt{\S}_n$ separates $\wt{V}$ into two noncompact ends and the open region between $\wt{\Sigma}_n$ and $\wt{\Sigma}_{n+1}$ is isometric to $V \setminus \S$. See e.g.,\  the proofs of \cite[Thm. 4.3]{AndGal} or \cite[Prop. 5]{GalLingCosmo}. From this we see that  $\wt{V}$ admits a natural $\Z$ action, $\wt{x} \mapsto n(\wt{x})$.
By introducing a time function on $M$ which lifts via the covering map to a time function on $\wt{M}$, we can assume $M = \R \times V$ and $\wt{M} = \R \times \wt{V}$ with $V$ and $\wt{V}$ embedding as the $t = 0$ slices. The $\Z$ action on $\wt{V}$  yields a corresponding $\Z$ action on $\wt{M}$ via $(t, \wt{x}) \mapsto \big(t, n(\wt{x})\big)$.
Applying Theorem \ref{thm: local rigidity 1} to the spacetime $\wt{M}$, it follows that each outward null expansion on $\wt{\S}_n$ vanishes. Let $W$ be the compact region between $\wt{\S}_0$ and $\wt{\S}_1$. Theorem \ref{thm: glob rigidity 1} applies and so $W$ is foliated by MOTS $\wt{\Sigma}_s$ for $s \in [0,1]$; moreover, each of these MOTS generates an achronal totally geodesic null hypersurface emanating from $\wt{\Sigma}_s$. The $\mathbb{Z}$ action on $\wt{V}$ and $\wt{M}$ along with a connectedness argument shows that $\wt{V}$ is foliated by  MOTS 
$\wt{\Sigma}_s$,  $s \in \R$, each diffeomorphic to $\S$ (so that $\wt{V}$ is topologically $\R \times \S$), and $\wt{M}$ is foliated by achronal totally geodesic null hypersurfaces emanating from $\wt{\Sigma}_s$ for $s \in \R$.

Let $\wt{N} \subset \wt{M}$ denote one of these null hypersurfaces. We claim that the covering map $\wt{M} \to M$ restricts to a one-to-one immersion $\wt{N} \to N \subset M$, where $N$ is the image of $\wt{N}$ under the covering map. That it's an immersion is clear. To see that it's one-to-one, suppose there are distinct points $\wt{p}_1, \wt{p}_2 \in \wt{N}$ each of which gets mapped to the same point $p \in N$. Then $\wt{p}_1$ and $\wt{p}_2$ have the following form: $\wt{p}_i = \big(t, n_i(\wt{x})\big)$. Each $\wt{p}_i$ is the future endpoint of a null geodesic $\wt{\g}_i \colon [0,b_i] \to \wt{M}_i$ with both $\wt{\g}_1$ and $\wt{\g}_2$ emanating from some $\wt{\Sigma}_s$. Set $n = n_1 - n_2$. The translated null geodesic $n \circ \wt{\g}_2$ begins at $\wt{\S}_{n + s}$ and ends at $\wt{p}_1$. Thus, if $n \neq 0$, then the null hypersurface emanating from $\wt{\S}_{n +s}$ intersects the one emanating from $\wt{\S}_s$ which is a contradiction. Thus $n = 0$ and the map $\wt{N} \to N$ is one-to-one.

Strong causality shows that the image $N$ is in fact embedded. Moreover, any two null hypersurfaces $\wt{N}_s$ and $\wt{N}_{s'}$ with $s \neq s'$, in the foliation (where we may assume that $\wt{N}_{s'}$ is in the timelike past of $\wt{N}_s$) project down to disjoint null hypersurfaces $N_{s}$ and $N_{s'}$. Indeed, if $N_s$ and $N_{s'}$ intersect tangentially then they coincide by the null maximum principle, and if they intersect transversely, then covering space arguments show that $\wt{N}_{s'}$ enters the timelike future of $\wt{N}_s$, which is a contradiction. Then $\{N_s \mid s \in [0,1)\}$ yields the desired foliation in $M$. 
 \qed

\medskip
\medskip

The following theorem explores some rigidity aspects of the main result in \cite{GalLingCosmo}. However, this theorem applies in all spacetime dimensions $\geq 3$.

\medskip

\begin{thm}\label{thm: cpt with boundary 2}
Let $(M,g)$ be a future null geodesically complete globally hyperbolic spacetime satisfying the null energy condition. Let $V$ be a spacelike compact Cauchy hypersurface, and let $\S$ be a closed nonseparating two-sided hypersurface in $V$. Assume $\theta^+ \geq 0$ and $\theta^- \geq 0$. 
Then either
\begin{itemize}
\item[\emph{(a)}] $M$ is past null geodesically incomplete, or else,
\item[\emph{(b)}] $M$ is foliated by two traverse families of closed totally geodesic null hypersurfaces; moreover, their intersections cover $M$ by totally geodesic codimension \emph{2} spacelike submanifolds.
\end{itemize}
\end{thm}

\proof
Assume $M$ is past null geodesically complete.
Let $\wt{\S}_n$, $\wt{V}$, and $\wt{M}$ be as in the covering construction in the proof of Theorem  \ref{thm: cpt with boundary 1}.  If there is a point on $\S_1$ such that $\theta^+ > 0$, then we can deform $\S_1$ via null mean curvature flow in the outward direction to produce a surface that's inner trapped towards the past, which contradicts past null completeness. Therefore $\theta^+ = 0$ on $\S_1$. Likewise $\theta^- = 0$ on $\S_1$. As in the proof of Theorem \ref{thm: cpt with boundary 1}, it follows that $\wt{V} \approx \R \times \Sigma$ is foliated by a family of MOTS $\{\wt{\S}_t^+\}_{t \in \R}$ and a family of MITS $\{\wt{\S}_s^-\}_{s \in \R}$ such that $\wt{\S}_n = \wt{\S}^+_n = \wt{\S}^-_n$ for all $n \in \Z$. These surfaces generate achronal totally geodesic null hypersurfaces $\wt{N}^+_t$ and $\wt{N}^-_s$. From Theorem \ref{thm: cpt with boundary 1}, we obtain two foliations by totally geodesic null hypersurfaces $N^+_t$ and $N^-_s$ in $M$. 
 Let $N$ denote the union of the intersections $\{N^+_t \cap N^-_s \mid t,s \in [0,1) \}$. Each of these intersections is a totally geodesic spacelike submanifold. As in the proof of Theorem~\ref{thm: glob rigidity 2}, a connectedness argument shows that $N = M$.
\qed

\medskip
\medskip

\noindent\emph{Remarks.} Theorem \ref{thm: cpt with boundary 2} can be viewed as a cosmological singularity theorem: apart from the very special circumstances detailed in part (b), $(M,g)$ must be past null geodesically complete, as examples of FLRW models illustrate.

 The assumption $\theta^\pm \geq 0$ in Theorem \ref{thm: cpt with boundary 2} is realized if, for example, the second fundamental form $K$ on $V$ is positive semi-definite and if $V$ is orientable and has nontrivial second homology. Indeed, by well-known arguments from geometric measure theory, the latter assumption implies that there is a closed nonseparating two-sided minimal hypersurface $\S$ in $V$, and since $K$ is positive semi-definite and $\S$ is minimal, we have $\theta^\pm \geq 0$. 

%
A simple example of a geodesically complete spacetime where Theorem \ref{thm: cpt with boundary 2} holds is the 
product spacetime $M = \R \times V$ where $g = -dt^2 + h$ and $(V,h)$ is the flat $n$-torus, and $\Sigma \subset V$ is any one of its nonseparating ($n-1$)-torus hypersurfaces. These examples show that the null hypersurfaces in the conclusion of Theorems \ref{thm: cpt with boundary 1} and \ref{thm: cpt with boundary 2} are not necessarily achronal. Another example is the Nariai spacetime which is just the product of two-dimensional de Sitter spacetime and $S^2$ \cite{BousHawk, Griffiths_Exact_Solutions}. 

\section{Topological censorship rigidity}\label{sec: top cen}


Topological censorship theorems show that the topology of the region exterior to a black hole is restricted in various respects, see e.g.,
\cite{FSW, Gdoc, GSWW, LingLesourd}.  An important precursor to the principle of topological censorship (in conjunction with the weak cosmic censorship conjecture) is the Gannon-Lee singularity theorem 
(\cite[Prop.\ 1.2]{Gannon}, \cite{Lee}), which establishes conditions under which nontrivial topology leads to null geodesic incompleteness; see also \cite{Costa, CostaMinguzzi, Stein, Gal_Ling_Null}.  In this section we obtain some rigidity results in this context.

\begin{thm}\label{thm: top cen local}
Let $(M,g)$ be a past null geodesically complete globally hyperbolic spacetime satisfying the null energy condition. Let $\S$ be a connected and simply connected closed hypersurface in a spacelike Cauchy hypersurface $V$ which separates $V$ into an inner component $V^-$ and outer component $V^+$, whose closures are compact and noncompact, respectively.  Assume $\theta^- \leq 0$ with respect to the normal on $\Sigma$ pointing into $V^+$.

Suppose $\ov{V^-} = V^- \sqcup \S$ is not simply connected. Then either
\begin{itemize}
\item[\emph{(a)}] $M$ is future null geodesically incomplete, or else,
\item[\emph{(b)}]  there is a neighborhood $U$ of $\S$ in $V$ such that $U \approx (-\e, \e) \times \S$ and for each $t \in (-\e, \e)$,  $\S_t = \{t\} \times \S$ is a MITS (in fact $\chi^-_t = 0$) and the inward null normal geodesics to $\S_t$ form a closed totally geodesic  immersed null hypersurface.
\end{itemize}
\end{thm}

\proof
Suppose $M$ is future null geodesically complete. Set $E_1 = \ov{V^-}$ and $E_2 = \ov{V^+}$. Let $p \colon \wt{E}_1 \to E_1$ denote the universal covering of $E_1$. By assumption $\wt{E}_1$ contains more than one copy of $\wt{\S} := p^{-1}(\Sigma)$. Since $\S$ is simply connected, $\wt{\S}$ is the disjoint union of $k$ diffeomorphic copies of $\S$, where $k$ is the cardinality of $\pi_1(E_1)$. Attach $k$ copies of $E_2$ to each copy of $\S$ in $\wt{\S}$ (as they're attached in the base space). Call the resulting space $\wt{V}$, and, abusing notation, let $p \colon \wt{V} \to V$ denote the resulting covering. Using covering space theory (as outlined in the beginning of section 3.2 in \cite{Gal_Ling_Null}) or the Bernal-Sanchez splitting result (as outlined in \cite[Lemma 4]{GalLingCosmo}), there is a corresponding spacetime covering $P \colon \wt{M} \to M$ such that $\wt{V}$ is a Cauchy surface for $\wt{M}$ and $P|_{\wt{V}} = p$. Let $\hat{\S}$ denote one of the components of $\wt{\S}$. Since $k \geq 2$ and $E_2$ is noncompact, it follows that $\hat{\S}$ separates $\wt{V}$ into two components whose closures are both noncompact. Therefore we can apply Theorem \ref{thm: local rigidity 1} (with outward and inward directions reversed) to $\hat{\Sigma}$ in the covering spacetime and then project down to the physical spacetime to obtain the result.\qed

\medskip

The $\R\mathbb{P}^n$ cylinder example, defined after Corollary \ref{cor: top cen}, shows that `immersed' rather than `embedded' is needed in the conclusion of Theorem \ref{thm: top cen local}.

\medskip

\noindent\emph{Remark.} If we assumed $\Sigma$ was weakly trapped in Theorem \ref{thm: top cen local}, i.e., $\theta^+ \leq 0$ and $\theta^- \leq 0$, then the conclusion can be strengthened so that the neighborhood $U$ of $\S$ in $V$ is as in the conclusion of Theorem \ref{thm: local rigidity 2}.

\medskip

\begin{thm}\label{thm: top cen normal subgroup}
Let $(M,g)$ be a past null geodesically complete globally hyperbolic spacetime. Let $\S$ be a connected and simply connected closed hypersurface in a spacelike Cauchy hypersurface $V$ which separates $V$ into an inner component $V^-$ and an outer component $V^+$, whose closures are compact and noncompact, respectively. Assume $\Sigma$ is weakly inner trapped and outer untrapped, $\theta^- \leq 0$ and $\theta^+\geq 0$, with respect to the unit normal pointing into $V^+$.

If $\pi_1(\ov{V^-})$ contains a proper subgroup $H$ with finite index, then either

\begin{itemize}
\item[\emph{(a)}] $M$ is future null geodesically incomplete, or else,
\item[\emph{(b)}] $\ov{V^-}$ is double covered by $\S \times [0,1]$. 
\end{itemize}
\end{thm}

\proof
Suppose $M$ is future null geodesically complete. Let $k < \infty$ denote the finite index. There is a $k$-sheeted covering $\wt{E}_1 \to E_1$, where, as in the proof of Theorem~\ref{thm: top cen local}, we set $E_1 = \ov{V^-}$ and $E_2 = \ov{V^+}$. Again, extend this covering to a $k$-sheeted covering $p \colon \wt{V} \to V$ by attaching $k$ copies of $E_2$ to each copy of $\S$ in the cover $\wt{E}_1$ and consider the corresponding spacetime covering $P \colon \wt{M} \to M$, where $\wt{V}$ is a Cauchy surface for $\wt{M}$. Since $2 \leq k < \infty$, it follows that $\wt{E}_1$ is compact since $E_1$ is. Let $\hat{\S}$ denote one component of $\wt{\S} := p^{-1}(\Sigma)$ and set $S := \wt{\S} \setminus \hat{\Sigma}$. We have $\theta^-_{\hat{\Sigma}} \leq 0$ (with respect to the normal pointing out of $\wt{E}_1$) and $\theta^+_S \geq 0$ (with respect to the normal pointing out of $\wt{E}_1$). The result follows by applying Theorem \ref{thm: glob rigidity 1} to the covering spacetime $\wt{M}$ with $W$ playing the role of $\wt{E}_1$. \qed

\medskip
\medskip

 Clearly Theorem  \ref{thm: top cen normal subgroup} applies whenever $\pi_1(\ov{V^-})$ is finite (just take $H$ to be the identity). But Theorem \ref{thm: top cen normal subgroup} can also apply when the fundamental group of $\pi_1(\ov{V^-})$ is infinite, e.g., it applies when $\ov{V^-} \approx \mathbb{B}^n \# (\mathbb{S}^1 \times \mathbb{S}^n)$, where $\mathbb{B}^n$ is the closed $n$-ball.

\medskip
\medskip

\noindent\emph{Remark.} In Theorem \ref{thm: top cen normal subgroup} the existence of $H$ with finite index was assumed to obtain a finite covering, so we could apply Theorem \ref{thm: glob rigidity 1}. Suppose, as in the remark following  Theorem~\ref{thm: glob rigidity 1}, one assumes that $V$ is geodesically complete, with bounded geometry, and that $K$ is uniformly bounded on all of $V$.  (These are properties that lift to covers.) By making use of the MOTS convergence discussed in that remark, the proof of 
Theorem~\ref{thm: top cen normal subgroup}, with these additional assumptions on $V$, goes through in essentially the same way (with the covering being the universal one), and the only assumption on $\pi_1(\ov{V^-})$ is that it's nontrivial. 

\medskip
\medskip

\begin{cor}\label{cor: top cen}
Let $(M,g)$ be a past null geodesically complete globally hyperbolic spacetime with ${\rm dim}(M) = 4$. Let $\S$ be a topological two-sphere in a spacelike Cauchy hypersurface $V$ which separates $V$ into an inner component $V^-$ and an outer component $V^+$, whose closures are compact and noncompact, respectively. Assume $\Sigma$ is weakly inner trapped and outer untrapped, $\theta^- \leq 0$ and $\theta^+\geq 0$, with respect to the unit normal pointing into $V^+$. Then one of the following must hold:
\begin{itemize}
\item[\emph{(a)}] $M$ is future null geodesically incomplete,
\item[\emph{(b)}] $\ov{V^-}$ is topologically a closed three-ball $\mathbb{B}^3$,
\item[\emph{(c)}] $\ov{V^-}$ is topologically $\mathbb{RP}^3$ minus an open three-ball.
\end{itemize}
\end{cor}

\proof
Suppose $M$ is future null geodesically complete.

 Either $\ov{V^-}$ is simply connected or not. Assume the former. Since $\S \approx \mathbb{S}^2$, we can attach three-ball to $\ov{V^-}$ along $\Sigma$. The resulting space is closed and simply connected, hence it's a three-sphere by the positive resolution of the Poincar{\'e} conjecture. Therefore $\ov{V^-}$ is a three-ball by a theorem of Alexander.  

Now assume $\ov{V^-}$ is not simply connected. The work of Hempel \cite[Thm.\ 1.1]{Hempel_res_finite} in conjunction with the positive resolution of the geometrization conjecture shows that $\pi_1(\ov{V^-})$ is residually finite, i.e., for every non-identity element in the group, there is a normal subgroup with finite index that does not contain that element. (See the introduction of \cite{3_mfld_groups} for a discussion.) Theorem \ref{thm: top cen normal subgroup} implies that $\ov{V^-}$ is double covered by $\S \times [0,1]$. Again, attach a three-ball to $\ov{V^-}$ along $\S$; in the cover this corresponds to capping off both ends of $\S \times [0,1]$ by three-balls to obtain a three-sphere. Hence the resulting space is double covered by $\mathbb{S}^3$ and hence it's topologically $\mathbb{RP}^3$ by the positive resolution of the elliptization conjecture. The result follows.
\qed

\medskip
\medskip

\noindent\emph{Examples.} An example of a geodesically complete spacetime where Corollary \ref{cor: top cen}(b) holds is Minkowski spacetime $M$, where $V$ is any $t = \rm{constant}$ slice and $\Sigma$ is any round sphere in $V$. An example of a geodesically complete spacetime where Corollary \ref{cor: top cen}(c) holds -- in any dimension -- is the product spacetime $M = \R \times V$, where $V$ is the $\R \mathbb{P}^n$ cylinder and $\Sigma$ is any one of the minimal latitudinal spheres which encloses the one-sided $\R \mathbb{P}^{n-1}$ equator. Here, by ``$\R\mathbb{P}^n$ cylinder" we mean the natural $\mathbb{Z}_2$-identification $(r,\omega) \mapsto (-r, -\omega)$  of the Riemannian product space $\R \times \mathbb{S}^{n-1}$, where $r \in \R$ and $\omega \in \mathbb{S}^{n-1}$. Note that the $\R\mathbb{P}^n$ cylinder is topologically $\R\mathbb{P}^n$ minus a point.

\medskip
\medskip

Lastly, we remark on the case when $\S$ is not simply connected. This case was originally studied in \cite{Costa, CostaMinguzzi} and refined in \cite{Gal_Ling_Null}. In this case, analogous results hold in this section, but with more/different hypotheses. In particular, we always require the additional assumption, ``the inclusion map $\S \hookrightarrow \ov{V^+}$ on fundamental groups is an isomorphism." With this assumption, an analogous result of Theorem \ref{thm: top cen local} holds but the assumption that ``$\pi_1(\ov{V^-})$ is nontrivial" is replaced with the assumption ``the induced homomorphism of the inclusion map $\S \hookrightarrow \ov{V^-}$ on fundamental groups is not surjective." See section 3.1 in \cite{Gal_Ling_Null} for details.

Similarly, a result like Theorem \ref{thm: top cen normal subgroup} also holds. In this case, again one assumes the technical assumption in the above paragraph. The hypothesis ``$\pi_1(\ov{V^-})$ contains a proper subgroup $H$ with finite index" is replaced with the assumption ``the induced homomorphism of the inclusion map $\Sigma \hookrightarrow \ov{V^-}$ is not surjective and $H$ has finite index within $\pi_1(\ov{V^-})$, where $H$ is the image of $\pi_1(\Sigma)$ into $\pi_1(\ov{V^-})$ via the induced homomorphism of the inclusion map" and the conclusion is that $\pi_1(\ov{V^-}) / H = \mathbb{Z}_2$.
Also, a similar reasoning as in the remark after Theorem \ref{thm: top cen normal subgroup} applies.

\appendix

\section{Proof of Lemma \ref{lem: IVT argument}}

\noindent\emph{Proof of Lemma \ref{lem: IVT argument}.} For small $f \in C^\infty(\S)$, let $\Theta^+(f)$ denote the outward null expansion of the hypersurface $\S_f \colon x \to \exp_x\big(f(x) \nu\big)$ with respect to the (suitably normalized) future directed outward null normal field to $\S_u$. $\Theta^+$ has linearization, $\Theta^{+\prime}(0) = L$, where $L$ is the MOTS stability operator introduced in section \ref{sec: prelim}. We introduce the operator
\[
\Phi\colon C^\infty(\S) \times \R \to C^\infty(\Sigma) \times \R, \quad \Phi(f,k) \,=\, \left(\Theta^+(f) -k, \int_\Sigma f \right).
\]
Since $\l_1 = 0$ is a simple eigenvalue, the kernel of $\Theta^{+\prime}(0) = L$ consists only of constant multiples of the positive eigenfunction $\phi$. Moreover, $\l_1 = 0$ is a simple eigenvalue for the adjoint $L^*$ of $L$ \cite[Lem. 4.1(ii)]{AMS08} (with respect to the $L^2$ inner product on $\S$); let $\phi^*$ denote a corresponding positive eigenfunction. Then the equation $Lf = v$ is solvable if and only if $\int_\S v\phi^* = 0$ (by the Fredholm alternative for elliptic operators). The linearization of $\Phi$ is
\[
(f,k) \mapsto \left(Lf - k, \int_\Sigma f \right).
\]
We show $\Phi$ has invertible linearization about $(0,0)$. Injectivity: Suppose $Lf -k = 0$ and $\int_\S f = 0$; the former implies $\int_\Sigma k \phi^* = 0$ which implies $k =0$; hence $f$ is a multiple of $\phi$ and so the latter implies $f = 0$. Surjectivity: Fix a function $h \in C^\infty(\S)$ and $a \in \R$. We want to find a function $f$ and number $k$ such that $Lf - k = h$ and $\int_\S f = a$. $Lf = h+k$ implies $\int_\S \phi^*(h + k) = 0$; thus we must have $k = -(\int_\S \phi^*h)/(\int_\S \phi^*)$. There exists an $f_0$ such that $Lf_0 = h+k$. Consider $f = f_0 + c\phi$. Then $Lf = h+k$ also. Moreover, $\int_\S f = \int_\S f_0 + c \int_\S \phi$. Therefore choose $c$ such that $\int_\S f = a$. 

 Thus, by the inverse function theorem, for $t \in \R$ sufficiently small, there exists $f(t) \in C^\infty(\S)$ and $k(t) \in \R$ such that 
\[
\Theta^+\big(f(t)\big) \,=\, k(t) \quad \text{ and } \quad \int_\S f(t) \,=\, t.
\]
By the chain rule, we have $\Theta^{+\prime}(0)\big(f'(0)\big) = L\big(f'(0)\big) = k'(0)$. Therefore $k'(0)$ is orthogonal to $\phi^*$. Hence $k'(0) = 0$, and so $f'(0)$ is in the kernel of $L$. Thus $f'(0)$ is a constant multiple of $\phi$, and since $\int_\S f'(0) = 1$, it follows that $f'(0) > 0$. Moreover, by invertibility, we have $f(0) = 0$.

It follows that for $t$ sufficiently small, the hypersurfaces $\S_t := \S_{f(t)}$ form a smooth foliation of a neighborhood of $\S$ in $V$ by hypersurfaces of constant outward null expansion. Thus, one can introduce coordinates $(t,x^i)$ in a neighborhood $U$ of $\S$ in $V$ such that, with respect to these coordinates, $U = (-\e, \e) \times \S$, and for each $t \in (-\e, \e)$, the $t$-slice $\S_t = \{t\} \times S$ has constant outward null expansion $\theta^+(t)$ with respect to $\ell(t) = u + \nu(t)$ where $\nu$ is the outward unit normal field to $\S_t$ in $V$. Moreover, the coordinates $(t, x^i)$ can be chosen so that $\pd_t = \phi \nu$, for some positive function $\phi$ on $U$. It follows that the metric on $U$ is given by $h|_U = \phi^2 dt^2 + \g_t$ where $\g_t$ is the induced metric on $\S_{t}$. \qed

\section*{Acknowledgments} 
We're grateful to Michael Eichmair for many helpful comments on an earlier draft. Gregory Galloway was supported by  the Simons Foundation, Award No. 850541. Eric Ling was supported by Carlsberg Foundation CF21-0680 and Danmarks Grundforskningsfond CPH-GEOTOP-DNRF151.

\vspace{.1in}
\noindent
{\bf Declarations.} On behalf of both authors, the corresponding author states that there is no conflict of interest. No data was collected or analysed as part of this project.


\providecommand{\bysame}{\leavevmode\hbox to3em{\hrulefill}\thinspace}
\providecommand{\MR}{\relax\ifhmode\unskip\space\fi MR }
\providecommand{\MRhref}[2]{%
  \href{http://www.ams.org/mathscinet-getitem?mr=#1}{#2}
}
\providecommand{\href}[2]{#2}

\end{document}